\documentstyle[12pt,epic]{article}

\begin{document}

\begin{titlepage}

\title{On the Algebraic Structure of the Holomorphic
Anomaly for $N=2$ Topological Strings}

\author{C\'esar G\'omez\thanks{Departement de Physique
Theorique, Universit\'e de Gen\`eve.} \thanks{Instituto
de Matem\'aticas y F\'isica Fundamental, Serrano 123,
28006 Madrid (Spain).}
 and Esperanza
L\'opez$^{\mbox{\footnotesize{\dag}}}$
\\ Th-Division, CERN \\ 1211 Gen\`eve 23, Switzerland }

\date{}

\maketitle

\vspace{8mm}

\begin{abstract}
The special geometry ($(t, {\bar t})$-equations) for twisted $N
\! = \! 2$ strings are derived as consistency conditions of a
new contact term algebra.
The dilaton field appears in the contact
terms of topological and antitopological operators.
The holomorphic anomaly, which can be
interpreted as measuring
the background dependence, is obtained from the contact
algebra relations.
\end{abstract}

\vspace{-13.5cm} \hspace{12cm} CERN-TH.7258/94

\end{titlepage}

\subsection*{Introduction}

For $N=2$ twisted topological strings, the problem of background
independence seems "ab initio" an almost trivial issue. In fact
once we fix a background by choosing a particular point
$(t_0,\bar{t}_{0})$ in the moduli space $\cal M$ of
the underlying $N=2$ SCFT, a
change in the background turns out to be equivalent, for the
twisted theory, to the coupling of a pure BRST state, which by
standard procedures should be expected to vanish. The
holomorphic anomaly recently discovered in reference\cite{BCOV1,BCOV2},
clearly shows
that this naive picture drastically changes after
coupling to gravity. As we will prove in this letter what makes
the holomorphic anomaly different from the standard BRST
anomaly is its special nature as a string amplitude. In fact the
holomorphic anomaly can be interpreted as defining a new
type of string amplitudes where the topological and
antitopological sectors are explicitly fused. In this letter we
will reinterpret these $(t,\bar{t})$ amplitudes in terms of a contact
term algebra, similar to the one defining pure topological
gravity \cite{witten1,VV}. The consistency conditions of this algebra
are proved to be the special geometry relations \cite{SG}
for the moduli space of the $N=2$ SCFT \cite{BCOV2}.
In reference\cite{BCOV2} an integrated version of the holomorphic
anomaly is defined in terms of a particular set of Feynman
rules. The most surprising fact about these rules is the
appearance of a dilaton field, defined by the same recursion
relations that in topological gravity \cite{witten1,VV}.
We will show that this
dilaton field is crucial for defining a fused
$(t,\bar{t})$  contact term algebra which completely determines the
$(t,\bar{t})$ amplitudes.

\subsection*{$(t,\bar{t})$-amplitudes and moduli measures}

For  $N \! = \! 2$ twisted SCFT's with $ \hat{c} \! = \! 3$,
the coupling to gravity is
obtained by defining the following measures \cite{of} on moduli space of
Riemann surfaces:

i) Partition function measures:
The measures on the moduli space ${\cal M}_{g}$ of Riemann
surfaces with genus $g$ are given by:

\begin{equation}
\int_{{\cal M}_g} \!  [dm] \; \langle \; \prod_{a,{\bar a}=1}^{3g-3}
\: G^{-}(\chi_a)
\bar{G}^{-}(\bar{\chi}_{\bar a}) \; \rangle_{\Sigma_g}
\label1
\end{equation}
\noindent
where the $\chi_{a}, {\bar \chi}_{\bar a}$ represent a basis
of Beltrami differentials.
For simplicity, we will adopt the notation referring to the A-twist, as
already done in (1); the corresponding expressions for the
B-twist can be immediately obtained replacing ${\bar G}^{-}$
by ${\bar G}^{+}$.

ii) Amplitude measures for truly marginal operators:
\begin{equation}
\int_{{\cal M}_{g,N}}  \! \! [dm] [dz]
\; \langle \; \prod_{i=1}^{N} \: \oint_{C_{z_i}}  \! \!
G^{-} \bar{G}^{-} \phi_{i} (z_i) \; \prod_{a,{\bar a}=1}^{3g-3}
\: G^{-}(\chi_a)
\bar{G}^{-} (\bar{\chi}_{\bar a}) \; \rangle_{\Sigma_{g,N}}
\label2
\end{equation}
\noindent
which define measures on the moduli space ${\cal M}_{g,N}$ of
Riemann surfaces with genus  $g$ and $N$ punctures.
Notice that only for marginal fields, which have $U(1)$
charge equal one, these measures are non
vanishing. In fact the $U(1)$ anomaly is completely saturated by
the $(3g-3)$ insertions of the supersymmetric $G^{-}$ charges.

The philosophy underlying these definitions is the formal
similarity between the twisted $N=2$ theory \cite{B,BLNW} and the bosonic
string:
\begin{eqnarray}
2j_{BRST} & \leftrightarrow & G^{+} \nonumber \\
b & \leftrightarrow & G^{-} \\
bc & \leftrightarrow & J \nonumber
\end{eqnarray}
\noindent
For the particular case ${\hat c} \! = \! 3$ the $U(1)$ anomaly
coincides with the dimension of the moduli space, making possible
to use the supersymmetric charges $G^{-}$ as the standard
$b$-ghost for the definition of the measure on the moduli space.

The background dependence of these amplitudes \cite{witten2}, i.e their
derivatives with respect to ${\bar t}_i$, are given by the
following "$(t,\bar{t})$ amplitude":
\begin{equation}
\int_{{\cal M}_{g,N}} \! \! [dm] [dz]  \int dw \; \langle \;
\oint_{C_w}  \! \! G^{+} \bar{G}^{+} \bar{\phi}_{\bar i}(w) \:
\prod_{i=1}^{N} \: \oint_{C_{z_i}} \! \! G^{-}
\bar{G}^{-} \phi_{i} (z_i) \: \prod_{a,{\bar a}=1}^{3g-3} G^{-}(\chi_a)
\bar{G}^{-} (\bar{\chi}_{\bar a}) \; \rangle
\end{equation}

\noindent
The string interpretation of this amplitude as a measure on the
corresponding moduli space ${\cal M}_{g,n+1}$, is by no means
direct. It is precisely at this point where the holomorphic
anomaly differs from the standard BRST-anomaly of the
bosonic string. In fact the integration of the moduli parameter
associated with the insertion of the antitopological field is
performed using the "antitopological ghost", i.e the supersymmetric
current $G^{+}$.
Searching for a standard string interpretation of the $(t,{\bar t})$
amplitude (4),
we can
formally try to represent it as follows:
\begin{equation}
\int_{{\cal M}_{g,N+1}} \! \! \! \! [dm] [dz] \; \langle \;
\oint_{C_w} \! \! G^{-} \bar{G}^{-} {\cal O}_{\bar i}(w)
\prod_{i=1}^{N} \: \oint_{C_{z_i}} \! \! G^{-}
\bar{G}^{-} \: \phi_{i} (z_i) \; \prod_{a,{\bar a}
=1}^{3g-3} \: G^{-}(\chi_a)
\bar{G}^{-} (\bar{\chi}_{\bar a}) \; \rangle_{\Sigma_{g,N+1}}
\end{equation}
\noindent
where we have introduced new operators ${\cal O}_{\bar i}$ with
positive ghost number equal one.
In order to make explicit the ghost number counting we can use
the following "gravitational picture".
We interpret the
operators ${\cal O}_{{\bar i}(i)}$ as having implicitely
a gravitational
descendent index $n$, i.e ${\cal O}_{{\bar i}(i),n}$. By
analogy with  topological matter coupled to topological gravity
models \cite{witten1,L},
their ghost number will be defined by the rule:
\begin{equation}
gh({\cal O}_{{\bar i}(i),n}) \: = \: n \: + \: q_{{\bar i}(i)}
\label6
\end{equation}
\noindent
with $q_{{\bar i}(i)}$ the $U(1)$ charge of the corresponding fields.
Following this prescription, the operators ${\cal O}_{\bar i}$
have $n=2$.
With the same logic we shall associate with the truly marginal
fields $\phi_i$, also of ghost number one, operators of the type
${\cal O}_{i,0}$. Summarizing the "gravitational picture" for
the $(t, \bar t)$ amplitudes is defined by the following set of formal
rules:
\begin{eqnarray}
\oint G^{-} {\bar G}^{-} \phi_{i} & \rightarrow &
\oint G^{-} {\bar G}^{-} {\cal O}_{i,0} \\
\oint G^{+} {\bar G}^{+} {\bar \phi}_{\bar i} &
\rightarrow & \oint G^{-} {\bar G}^{-}
{\cal O}_{\bar{i},2} \nonumber
\end{eqnarray}
\noindent
The philosophy underlying (7.b) can be thought as a process
in two steps. Inspirated by the relation between gravitational
descendants and pure BRST states in ${\hat c} \! < \! 1$
Landau-Ginzburg models coupled to gravity \cite{Lo,O}, we
first interpret the pure BRST insertion of the antitopological
field as representing a gravitational descendant. Secondly, we
integrate over the position of the corresponding puncture in the
standard topological way, namely using $G^{-}$ as the b-ghost.

One more reason supporting the
gravitational picture comes from the subleading
divergences in the
operator product:
\begin{equation}
{\bar \phi}_{\bar j}(z) \; \phi_{i}(0) \; =
\frac{ G_{i {\bar j}}}{|z|^2}
\end{equation}
\noindent
with $G_{i {\bar j}}$ the Zamolodchikov metric associated with
the truly marginal deformations \cite{Z}. These subleading  singularities
\cite{K}, which depend linearly on the curvature of the surface,
determine the "dilaton" contribution to the holomorphic anomaly
\cite{BCOV2}. The natural translation of them
into the gravitational
picture presented above,
would be defined by the following contact term:
\begin{equation}
\int_{D} {\cal O}_{{\bar j},2} | {\cal O}_{i,0} \rangle \: = \:
G_{i {\bar j}} \: | \sigma_{1} \rangle
\end{equation}
\noindent
with $\sigma_{1}$ representing a dilaton field, in the sense of
topological gravity, and the domain of integration D is an
infinitesimal neighborhood of the point where the field
${\cal O}_{i,0}$ is inserted.
The dynamics of the dilaton field will be later
determined by its contact terms with the rest of the fields.

The previous arguments should be interpreted only as providing
some heuristic
support for the rules defined in equation (7). Our strategy now
will be to find a contact term algebra for the operators in the
gravitational picture.
{}From this contact term
algebra we will be able to derive the holomorphic anomaly
and the special geometry of the moduli space of $N=2$ SCFT
($(t,\bar{t})$- equations \cite{CV}),
which will appear in this context
as the consistency conditions of the algebra.

\subsection*{The $(t,\bar{t})$ equations as consistency conditions
of a contact term algebra}

Let us consider the algebra of operators generated by:
${\cal O}_{i,0}$, ${\cal O}_{{\bar i},2}$ and the dilaton field
$\sigma_{1}$ with $i=1,..,n$ for $n$ the number of truly marginal
deformations. We define the following contact term algebra:
\begin{eqnarray}
\int_{D} {\cal O}_{i,0} \: | {\cal O}_{j,0} \rangle \: = \:
\Gamma_{i j}^{k} | {\cal O}_{k,0} \rangle
& , & \int_{D} {\cal O}_{{\bar i},2} | {\cal O}_{{\bar j},2}
\rangle \: = \:
{\tilde \Gamma}_{{\bar i} {\bar j}}^{\bar k}
| {\cal O}_{{\bar k},2} \rangle \nonumber \\
\int_{D} {\cal O}_{{\bar i},2} | {\cal O}_{j,0} \rangle \: = \:
G_{j {\bar i}} | \sigma_{1} \rangle
& , & \int_{D} {\cal O}_{i,0} | {\cal O}_{{\bar j},2} \rangle \: =  \:
{\widetilde G}_{i {\bar j}} | {\sigma_{1}} \rangle  \nonumber \\
\int_{D} {\sigma}_{1} \: | {\cal O}_{i,0} \rangle \: =  \:
a \:  | {\cal O}_{i,0} \rangle
& , & \int_{D} {\cal O}_{i,0} | {\sigma}_{1} \rangle \: =  \:
b | {\cal O}_{i,0} \rangle \\
\int_{D} \sigma_{1} | {\cal O}_{{\bar i},2} \rangle \: =  \:
c | {\cal O}_{{\bar i},2} \rangle
& , & \int_{D} {\cal O}_{{\bar i},2} | {\sigma}_{1} \rangle \: =  \:
d | {\cal O}_{{\bar i},2} \rangle \nonumber \\
\int_{D} {\sigma}_{1} | {\sigma}_{1} \rangle \: =  \:
e | {\sigma_{1}} \rangle && \nonumber
\end{eqnarray}
\noindent
In order to take into account the contribution of the curvature
we introduce the operator $e^{\frac{1}{2} \phi (z)}$,
the exponential of the bosonized $U(1)$ current.
The contact term algebra for this operator is
defined as follows:
\begin{eqnarray}
\int_{D} {\cal O}_{i,0} | e^{\frac{1}{2} \phi (z)} \rangle \: =  \:
A_{i}  | e^{\frac{1}{2} \phi (z)} \rangle
& , & \int_{D} {\cal O}_{{\bar i},2}
| e^{\frac{1}{2} \phi (z)} \rangle \: =  \: 0 \\
\int_{D} {\sigma}_{1} | e^{\frac{1}{2} \phi (z)} \rangle \: =  \:
a | e^{\frac{1}{2} \phi (z)} \rangle &  & \nonumber
\end{eqnarray}
\noindent

The undetermined constants appearing in (10) and (11) will be now
fixed by imposing the following consistency conditions:
\begin{equation}
\int_{D} a  \int_{D} b \; | c \rangle \: = \: \int_{D} b
\int_{D} a \; | c \rangle
\end{equation}
\noindent
for three arbitrary operators.
These conditions simply imply that the
string amplitudes are independent of the order in which the
fields are integrated\footnote{Notice that there are two
contributions to each side of (12): the successive contact terms of
the operators $a$ and $b$ with the bracketed $c$, and the
contact term between
$a$ and $b$ first, then carried over $| c \rangle $.}.
To solve the consistency conditions we
will assume:

i) That $G_{i,\bar j}$ is invertible.

ii) The value of $a$ equal -1. This condition is based on the
way the dilaton field measures the curvature.

iii) The following derivation rules:
\begin{eqnarray}
{\cal O}_{i,0} \Gamma_{\alpha \beta}^{\gamma}(t,{\bar t})
& = & \partial_{i} \Gamma_{\alpha \beta}^{\gamma}(t,{\bar t})  \\
{\cal O}_{{\bar i},2} \Gamma_{\alpha \beta}^{\gamma}(t,{\bar t})
& = &  (-1)^{F( \Gamma_{\alpha \beta}^{\gamma})} \;
\partial_{\bar i} \Gamma_{\alpha \beta}^{\gamma}(t,{\bar t})
\hspace {5mm}, \hspace{8mm}
F(\Gamma_{\alpha \beta}^{\gamma}) \! =  \! q_{\gamma} \!
- \! q_{\alpha}  \! - \! q_{\beta} \nonumber
\end{eqnarray}
\noindent
where $\Gamma_{\alpha \beta}^{\gamma}$ stands for a generic
contact term tensor, $q_{\alpha}$ for the $U(1)$ charge
associated to the field ${\cal O}_{\alpha}$, and
which defines the way the operators act on the coefficients
appearing in the contact term algebra. Notice that in general these
coefficients will depend on the moduli parameters $(t,{\bar t})$.
The logic for this rule is the equivalence
between the insertion of a marginal field and the derivation with
respect to the corresponding moduli parameter. For this reason
we will not associate any derivative with the dilaton field.
The derivation rule (13.b) is forced by the gravitational
picture we are using. Once we decide to work with gravitational
descendent and to define the measure using only $G^{-}$
insertions, we must accommodate to this picture the coupling
of the spin connection to the $U(1)$ current. Since the
derivation $\partial_{\bar i}$ correspond to the insertion
of an antitopological field,
in order to pass to the gravitational picture,sssss we need
to change, in the neighborhood of the insertion, the sign of
the coupling of the $U(1)$ current to the background gauge
field defined by the spin connection. This fact gives raise
to the factor $(-1)^{F(\Gamma)}$ in (13.b).

We will pass now, using i), ii) and iii), to solve the consistency
conditions (12). Let us start analyzing the following relation:
\begin{equation}
\int_{D} \sigma_{1} \int_{D} {\cal O}_{i,0} | {\cal O}_{j,0}
\rangle = \int_{D} {\cal O}_{i,0} \int_{D} \sigma_{1}
| {\cal O}_{j,0} \rangle \\
\end{equation}
\noindent
Applying the contact term algebra (10), we get:
\begin{equation}
b \: \Gamma_{i j}^{k} | {\cal O}_{k,0} \rangle -
\Gamma_{i j}^{k} | {\cal O}_{k,0} \rangle =
-2 \: \Gamma_{i j}^{k} | {\cal O}_{k,0} \rangle \\
\end{equation}
\noindent
which, for a non vanishing $\Gamma_{i j}^{k}$, implies that:
\begin{equation}
b= -1 \\
\end{equation}
\noindent
{}From the condition:
\begin{equation}
\int_{D} \sigma_{1} \int_{D} {\cal O}_{i,0} | \sigma_{1}
\rangle = \int_{D} {\cal O}_{i,0} \int_{D} \sigma_{1}
| \sigma_{1} \rangle \\
\end{equation}
\noindent
together with equation (16) and the derivation rules (13), we obtain:
\begin{equation}
\partial_{i} e | \sigma_{1} \rangle - e | {\cal O}_{i,0}
\rangle = | {\cal O}_{i,0} \rangle \\
\end{equation}
\noindent
being solved by:
\begin{equation}
e=-1 \\
\end{equation}
\noindent
To continue the study, we take the condition:
\begin{equation}
\int_{D} {\cal O}_{{\bar i},2} \int_{D} {\cal O}_{{\bar j},2}
| {\cal O}_{k,0} \rangle = \int_{D} {\cal O}_{{\bar j},2}
\int_{D} {\cal O}_{{\bar i},2} | {\cal O}_{k,0} \rangle \\
\end{equation}
\noindent
which leads to:
\begin{equation}
( \: {\tilde \Gamma}_{{\bar j} {\bar i}}^{\bar l} \: G_{k {\bar l}}
+  \partial_{\bar i} G_{k {\bar j}} \: ) \: | \sigma_{1} \rangle +
d \: G_{k {\bar j}} \: | {\cal O}_{{\bar i},2} \rangle =
(\: {\tilde \Gamma}_{{\bar i} {\bar j}}^{\bar l} \: G_{k {\bar l}}
+  \partial_{\bar j} G_{k {\bar i}} \: ) \: | \sigma_{1} \rangle +
d \: G_{k {\bar i}} \: | {\cal O}_{{\bar j},2} \rangle  \\
\end{equation}
\noindent
Using that $G_{i {\bar j}}$ is invertible, and for a general
number of truly marginal deformations, we get from the above equation:
\begin{equation}
d=0 \\
\end{equation}
\noindent
Moreover, the consistency condition:
\begin{equation}
\int_{D} \sigma_{1} \int_{D} {\cal O}_{{\bar i},2} | {\cal O}_{i,0}
\rangle = \int_{D} {\cal O}_{{\bar i},2} \int_{D} \sigma_{1}
| {\cal O}_{i,0} \rangle \\
\end{equation}
\noindent
and equation (22) imply that:
\begin{equation}
c=0 \\
\end{equation}
{}From (16), (19), (22) and the consistency condition:
\begin{equation}
\int_{D} {\cal O}_{i,0} \int_{D} {\cal O}_{{\bar j},2}
| \sigma_{1} \rangle = \int_{D} {\cal O}_{{\bar j},2}
\int_{D} {\cal O}_{i,0} | \sigma_{1} \rangle \\
\end{equation}
\noindent
we get easily:
\begin{equation}
{\widetilde G}_{i {\bar j}} = 0 \\
\end{equation}
\noindent
The next conditions we will analyze involve the curvature
operator $e^{\frac{1}{2} \phi (z)}$:
\begin{eqnarray}
\int_{D} {\cal O}_{i,0} \int_{D} {\cal O}_{{\bar j},2}
| e^{\frac{1}{2} \phi (z)} \rangle & = & \int_{D} {\cal O}_{{\bar j},2}
\int_{D} {\cal O}_{i,0} | e^{\frac{1}{2} \phi (z)} \rangle \\
\int_{D} {\cal O}_{i,0} \int_{D} {\cal O}_{j,0}
| e^{\frac{1}{2} \phi (z)} \rangle & = & \int_{D} {\cal O}_{j,0}
\int_{D} {\cal O}_{i,0} | e^{\frac{1}{2} \phi (z)} \rangle \nonumber
\end{eqnarray}
\noindent
from which we get, assuming that $\Gamma_{i j}^{k}$ is symmetric
in the lower
indices\footnote{The symmetry of $\Gamma_{i j}^{k}$ will assure that
$\int_{D} {\cal O}_{i,0} \int_{D} {\cal O}_{j,0}
| {\cal O}_{k,0} \rangle = \int_{D} {\cal O}_{j,0}
\int_{D} {\cal O}_{i,0} | {\cal O}_{k,0} \rangle$ is satisfied.}:
\begin{eqnarray}
 G_{i {\bar j}} & = &
 \partial_{\bar j} A_{i} \\
\partial_{i} A_{j} & =  & \partial_{j} A_{i} \nonumber
\end{eqnarray}
\noindent
Equations (28) imply that the metric
$G_{i {\bar j}}$ is K\"ahler, for a certain potential $K(t,{\bar t})$:
\begin{equation}
G_{i {\bar j}} = \partial_{i} \partial_{\bar j} K \\
\end{equation}
\noindent
With this information, we can return to (21) and deduce that the tensor
${\tilde \Gamma}_{{\bar i} {\bar j}}^{\bar k}$ is symmetric in
the lower indices:
\begin{equation}
{\tilde \Gamma}_{{\bar i} {\bar j}}^{\bar k} =
{\tilde \Gamma}_{{\bar j} {\bar i}}^{\bar k} \\
\end{equation}
\noindent
Using now:
\begin{equation}
\int_{D} {\cal O}_{i,0} \int_{D} {\cal O}_{{\bar j},2}
| {\cal O}_{{\bar k},2} \rangle  =
\int_{D} {\cal O}_{{\bar j},2} \int_{D} {\cal O}_{i,0}
| {\cal O}_{{\bar k},2} \rangle \\
\end{equation}
\noindent
we obtain that ${\tilde \Gamma}_{{\bar j} {\bar k}}^{\bar l}$
is only function of
the antitopological variables:
\begin{equation}
\partial_{i} {\tilde \Gamma}_{{\bar j} {\bar k}}^{\bar l} = 0 \\
\end{equation}
\noindent
Condition (32),
together with $\int_{D} {\cal O}_{{\bar i},2} \int_{D}
{\cal O}_{{\bar j},2}
| {\cal O}_{{\bar k},2} \rangle  =
\int_{D} {\cal O}_{{\bar j},2} \int_{D} {\cal O}_{{\bar i},2}
| {\cal O}_{{\bar k},2} \rangle$
allow to impose a vanishing contact term for antitopological operators.

To conclude the study of the consistency conditions we will
consider now the relation:
\begin{equation}
\int_{D} {\cal O}_{i,0} \int_{D} {\cal O}_{{\bar j},2}
| {\cal O}_{k,0} \rangle = \int_{D} {\cal O}_{{\bar j},2}
\int_{D} {\cal O}_{i,0} | {\cal O}_{k,0} \rangle \\
\end{equation}
\noindent
Using equations (16) and (26), we obtain:
\begin{eqnarray}
\int_{D} {\cal O}_{i,0} \int_{D} {\cal O}_{{\bar j},2}
| {\cal O}_{k,0} \rangle & = &
\partial_{i} G_{k {\bar j}} \: | \sigma_{1} \rangle -
G_{k {\bar j}} \: | {\cal O}_{i,0} \rangle - G_{i {\bar j}}
\: | {\cal O}_{k,0} \rangle  + fact \; terms \\
\int_{D} {\cal O}_{{\bar j},2}
\int_{D} {\cal O}_{i,0} | {\cal O}_{k,0} \rangle & = &
- \partial_{\bar j} \Gamma_{i k}^{l} \: | {\cal O}_{l,0} \rangle
+ \Gamma_{i k}^{l} G_{l {\bar j}} | \sigma_{1} \rangle \nonumber
\end{eqnarray}
\noindent
In topological gravity, the gravitational descendent
index required to factorize the surface, is $n \geq 2$. Therefore,
and due
to the non vanishing correlation
function $C_{i j}^{k}$ at genus zero for three marginal fields,
we should consider the existence of factorization terms associated to the
${\cal O}_{{\bar j}, 2}$ insertions.
We can write generically the factorization term as follows:
\begin{equation}
fact \; terms = B_{\bar j}^{l n} \, C_{i k n} \: | {\cal O}_{l,0}
\rangle \\
\end{equation}
\noindent
{}From equations (33)-(35),
we obtain that the
coefficient $\Gamma_{i j}^{k}$ is the connection for the metric
$G_{i {\bar j}}$, which we already know that is K\"ahler:
\begin{equation}
\Gamma_{i j}^{k} = ( \partial_{i} G_{j {\bar l}} )
G^{{\bar l} k} \\
\end{equation}
\noindent
and a $(t,{\bar t})$ type equation:
\begin{equation}
\partial_{\bar n} \Gamma_{i j}^{k} = G_{i {\bar n}} \delta_{j}^{k}
+ G_{j {\bar n}} \delta_{i}^{k} -
B_{\bar n}^{m k} C_{i j m} \\
\end{equation}

The tensor $B_{\bar j}^{l n}$ can be derived from the contact
term algebra by the following argument.
Let's consider the consistency condition on a general
string amplitude:
\begin{equation}
\langle {\cal O}_{{\bar i},2} {\cal O}_{{\bar j},2}
\prod_{l=1}^{N} {\cal O}_{l,0} \rangle_{g} =
\langle {\cal O}_{{\bar j},2} {\cal O}_{{\bar i},2}
\prod_{l=1}^{N} {\cal O}_{l,0} \rangle_{g} \\
\end{equation}
\noindent
from (10) we get:
\begin{equation}
(\, \Gamma_{{\bar i} {\bar j}}^{\bar k} -
\Gamma_{{\bar j} {\bar i}}^{\bar k} \, ) \langle \,
{\cal O}_{{\bar k},2} \,
\prod_{l=1}^{N} {\cal O}_{l,0} \, \rangle_{g} = \sum_{l=1}^{N}
{\cal R}_{D_{l}} + \sum_{nodes} {\cal R}_{\Delta} \\
\end{equation}
\noindent
where ${\cal R}_{D_{l}}$ denotes the
commutator of the contact terms of
${\cal O}_{{\bar i},2}$ and ${\cal O}_{{\bar j},2}$ with
${\cal O}_{l,0}$, and ${\cal R}_{\Delta}$ the commutator of those
at the nodes.
Using now the symmetry of
$\Gamma_{{\bar i} {\bar j}}^{\bar k}$
in the lower indices (27), we can conclude:
\begin{equation}
\sum_{l=1}^{N} {\cal R}_{D_{l}} = \sum_{nodes} {\cal R}_{\Delta} = 0
\end{equation}

The contribution at a node associated with the factorization of the
surface, will be
defined by the tensor $B_{\bar j}^{\alpha \beta}$ as follows \cite{VV}:
\begin{eqnarray}
\langle \, {\cal O}_{{\bar i},2} {\cal O}_{{\bar j},2}
\, \prod_{l \in S} {\cal O}_{l,0} \, \rangle_{g,{\Delta}}  & = &
\sum_{r=0}^{g} \:
\sum_{X \cup Y = S} [ \:
B_{\bar j}^{\alpha \beta} G_{\alpha {\bar i}}
\: \langle \sigma_{1}
\: \prod_{l \in X} {\cal O}_{l,0} \rangle_{r} \:
\langle {\cal O}_{\beta , 0} \: \prod_{n \in Y}
{\cal O}_{n,0} \rangle_{g-r} +  \nonumber \\
& + & \partial_{\bar i} B_{\bar j}^{\alpha \beta} \:
\langle \, {\cal O}_{\alpha ,0}
\prod_{l \in X} {\cal O}_{l,0} \, \rangle_{r}  \:
\langle {\cal O}_{\beta , 0} \: \prod_{n \in Y}
{\cal O}_{n,0} \rangle_{g-r} \: ]
\end{eqnarray}
\noindent
where $S$ refers to the set of all punctures, $X$ and $Y$
is a partition of it, and the tensor $B$ can be chosen symmetric
in the upper indices.
Using now (40) we get:
\begin{eqnarray}
B_{\bar i}^{\alpha \beta} G_{\alpha {\bar j}} & = &
B_{\bar j}^{\alpha \beta} G_{\alpha {\bar i}}  \\
\partial_{\bar i} B_{\bar j}^{\alpha \beta}
& = & \partial_{\bar j} B_{\bar i}^{\alpha \beta} \nonumber
\end{eqnarray}
\noindent
By an analogous argument, we find from
condition (33) and for a general string amplitude:
\begin{equation}
\partial_{i} B_{\bar j}^{\alpha \beta} + B_{\bar j}^{\alpha \gamma}
\Gamma_{i \gamma}^{\beta} + B_{\bar j}^{\gamma \beta}
\Gamma_{i \gamma}^{\alpha} - 2 \partial_{i} K
B_{\bar j}^{\alpha \beta} = 0
\end{equation}
\noindent
Let's define
$B_{\bar j}^{\alpha \beta} = B_{{\bar j} {\bar \alpha} {\bar \beta}}
e^{2K} G^{{\bar \alpha} {\alpha}} G^{{\bar \beta} {\beta}}$. Then,
equations (42) and (43) imply that $B_{{\bar i}{\bar j}{\bar \beta}}$
is proportional to the three point correlation function
for the antitopological fields.
Substituting this information into equation (37), we obtain
the $(t,{\bar t})$-equation \cite{CV}:
\begin{equation}
\partial_{\bar n} \Gamma_{i j}^{k} = G_{i {\bar n}} \delta_{j}^{k}
+ G_{j {\bar n}} \delta_{i}^{k} -
{\bar C}_{\bar n}^{m k} C_{i j m} \\
\end{equation}
\noindent
Notice that in order to get the special geometry relation (44)
from the contact term algebra, it was necessary to make use of
the derivation rule (13.b).
{}From (44) we can conclude that the metric
$G_{i {\bar j}}$ is the Zamolodchikov metric for the
marginal deformations.
With this result we finish the derivation of the $(t,{\bar t})$
equations as consistency conditions for the contact term algebra
(10). Our next objective will be the derivation of the
holomorphic anomaly.

\subsection*{The Holomorphic Anomaly}

The holomorphic anomaly
\cite{BCOV1,BCOV2}, i.e dependence on the background point,
is determined by the $(t,{\bar t})$-amplitudes (4). In this section
we will compute these amplitudes by using the gravity
representation defined by the formal rules (7) and
the machinery of the contact term algebra.
In the gravitational picture, the amplitude (4) becomes:
\begin{eqnarray}
& \partial_{\bar t} & C_{i_{1} \ldots i_{N}}^{g}  =  \nonumber \\
& = & \int_{{\cal M}_{g,N+1}} \! \! [dm] [dz] \; \langle \;
\oint_{C_z} \! G^{-} \bar{G}^{-} {\cal O}_{\bar t}(z)
\prod_{i=1}^{N} \: \oint_{C_{z_i}} \: G^{-}
\bar{G}^{-} \! \phi_{i} (z_i) \; \prod_{a,{\bar a }=1}^{3g-3}
\: G^{-}(\chi_a)
\bar{G}^{-} (\bar{\chi}_{\bar a}) \; \rangle_{\Sigma_{g,N+1}} =
\nonumber \\
& = & \langle \: {\cal O}_{{\bar t},2} \prod_{i=1}^{N}
{\cal O}_{i,0} \: \rangle_{g}
\end{eqnarray}
\noindent
where we have introduced the last equality to simplify the notation.
The contributions to (45) can be written:
\begin{equation}
\langle \: {\cal O}_{{\bar t},2} \prod_{i=1}^{N}
{\cal O}_{i,0} \: \rangle_{g} = \sum_{i=1}^{N} R_{D_{i}}
+ \sum_{nodes} R_{\Delta} \\
\end{equation}
\noindent
where $R_{D_{i}}$ is the contact term of
${\cal O}_{{\bar t},2}$ with the ${\cal O}_{i,0}$ insertion, and
$R_{\Delta}$ the contact term contribution that factorize the
surface through a node.
Let's start by analyzing the $R_{D_{i}}$ boundaries:
\begin{eqnarray}
\sum_{i=1}^{N} R_{D_{i}} & = & \sum_{i=1}^{N} \: \langle \:
{\cal O}_{{\bar t},2} \prod_{j=1}^{N} {\cal O}_{j,0} \:
\rangle_{D_{i}} \: = \: \sum_{i=1}^{N} \:
G_{i {\bar t}} \: \langle \: \sigma_{1} \prod_{j \neq i}
{\cal O}_{j,0} \: \rangle = \: \nonumber \\
& = & \sum_{i=1}^{N} \: G_{i {\bar t}} \:
(2 \! - \! 2g \! - \! n \! + \! 1) \: \langle
\: \prod_{j \neq i} {\cal O}_{j,0} \: \rangle
\end{eqnarray}
\noindent
The internal nodes $\Delta$ are associated to the
two types of boundaries of a Riemann surface of genus
$g$ and $N$ punctures.
The first one, we will note it as $\Delta_{1}$,
comes from pinching a handle, leading to a surface of
genus $g \! - \! 1$:
\begin{equation}
\langle \: {\cal O}_{{\bar t},2} \prod_{i=1}^{N} {\cal O}_{i,0} \:
\rangle_{g, \: {\Delta_{1}}} \; = \; \frac{1}{2} \:
B_{\bar t}^{' \alpha \beta}
\langle \: {\cal O}_{\alpha , 0} {\cal O}_{\beta ,0}
\prod_{i=1}^{N} {\cal O}_{i,0} \: \rangle_{g-1} \\
\end{equation}
\noindent
where the factor $\frac{1}{2}$ should be added
to reflect the equivalency between the order
in which the two new insertions ${\cal O}_{\alpha ,0}$
are integrated.
The factorization tensor $B'$ satisfies the same set of
equations (42) and (43) that the tensor $B$,
thus it is also proportional to the three point correlation function.
With an appropriate choice of normalization of the string amplitudes,
the proportionality constant between both factorization
tensors can be set equal to one \cite{L}.

The second ones, noted $\Delta_{2}$,
come from the factorization of the surface into
two surfaces of genus $r$ and $s$ punctures,
and genus $g-r$ and $N-s$ punctures
respectively:
\begin{equation}
\langle \: {\cal O}_{{\bar t},2} \prod_{i \in S} {\cal O}_{i,0} \:
\rangle_{g, \: {\Delta_{2}}} \; = \; \frac{1}{2} \: \sum_{r=0}^{g} \:
\sum_{X \cup Y = S}
B_{\bar t}^{\alpha \beta} \: \langle {\cal O}_{\alpha , 0}
\: \prod_{j \in X} {\cal O}_{j,0} \rangle_{r} \:
\langle {\cal O}_{\beta , 0} \: \prod_{k \in Y}
{\cal O}_{k,0} \rangle_{g-r} \\
\end{equation}
\noindent
Collecting now equations (47), (48) and (49), we obtain the
equation for the $\bar t$-dependence of any string amplitude:
\begin{eqnarray}
\partial_{\bar t} \: \langle \: \prod_{i \in S} {\cal O}_{i,0}
\: \rangle_{g}
& = & \frac{1}{2} \: B_{\bar t}^{\alpha \beta} \:
\langle \: {\cal O}_{\alpha , 0} {\cal O}_{\beta ,0}
\prod_{i \in S} {\cal O}_{i,0} \: \rangle_{g-1} + \nonumber \\
& + & \frac{1}{2} \: \sum_{r=0}^{g} \:
\sum_{X \cup Y = S}
B_{\bar t}^{\alpha \beta} \: \langle \: {\cal O}_{\alpha , 0}
\: \prod_{j \in X} {\cal O}_{j,0} \: \rangle_{r} \:
\langle \: {\cal O}_{\beta , 0} \: \prod_{k \in Y}
{\cal O}_{k,0} \: \rangle_{g-r} \; + \\
& + & \sum_{i \in S} \: G_{i {\bar t}}\:
(2 \! - \! 2g \! - \! n \! + \! 1) \: \langle
\: \prod_{j \neq i} {\cal O}_{j,0} \: \rangle_{g} \nonumber
\end{eqnarray}
\noindent
Notice that in our derivation of the holomorphic anomaly from
the contact term algebra we have only considered the contact
terms of the antitopological operator ${\cal O}_{{\bar t},2}$
with the rest of the operators ${\cal O}_{i,0}$ but not the
contact terms among the operators  ${\cal O}_{i,0}$ themselves.
This is equivalent to define the correlators
$\langle \prod {\cal O}_{i,0} \rangle$
by covariant derivatives of the generating functional. There are
however some aspects of the previous derivation that should be
stressed at this point.

1) The correlators $\langle \prod {\cal O}_{i,0} \rangle$
for topological operators can not be
determined by the contact term algebra, by contrast to what
happen in topological gravity. In fact from the contact
term algebra we can only get relations of the type:
\begin{equation}
(\, \Gamma_{i j}^{k} -
\Gamma_{j i}^{k} \, ) \langle \,
{\cal O}_{k,0} \,
\prod_{l=1}^{N} {\cal O}_{l,0} \, \rangle = \sum_{l=1}^{N}
{\cal R}_{D_{l}} + \sum_{nodes} {\cal R}_{\Delta} \\
\end{equation}
\noindent
which does not imply ($\Gamma_{i j}^{k} -
\Gamma_{j i}^{k} = 0$) anything on the surface
contribution. Moreover they are compatible with making all
contact terms $R_{D_{l}}$ equal to zero by covariantization.

2) If in the computation of
$\langle \, {\cal O}_{{\bar t},2} \,
\prod_{i=1}^{N} {\cal O}_{i,0} \, \rangle$
we take into account all
contact terms, i.e contact terms between the ${\cal O}_{i,0}$
operators, we
will find, as a consequence of the derivation rules (13) and the
$(t,{\bar t})$ equations (44), that the holomorphic anomaly is
cancelled, reflecting the commutativity of ordinary derivatives
$[ \partial_{\bar i} , \partial_{j} ] = 0$.

3) We should say that from the contact term algebra we can not prove,
at least directly, that the correlators
$\langle \, {\cal O}_{{\bar t},2} \,
\prod_{i=1}^{N} {\cal O}_{i,0} \, \rangle$
are saturated by
contact terms. The fact we have proved is that the contact
term contribution dictated by the contact term algebra (10) is
precisely the holomorphic anomaly.

4) The curvature of the initial surface is augmented by
two units in both
processes of pinching a handle or factorizing the surface. In order
to take this into account, the two insertions ${\cal O}_{\alpha
,0}, {\cal O}_{\beta , 0}$
generated in these processes should include, in addition, an
extra unit of curvature. Therefore, the total balance of curvature
for the new insertions is zero. This can be seen as the reason for
the zero contact term between the dilaton field $\sigma_{1}$
and the antitopological operators ${\cal O}_{{\bar i},2}$ (see
equations (28) and (30)).

\subsection*{Final Comments}

The main result of this letter was the derivation of the
$(t,{\bar t})$-equations as consistency conditions of a contact
term algebra. Using this point of view we get a more direct
understanding on the connections between the
$(t,{\bar t})$-equations,
the holomorphic anomaly and the dynamical role of the
dilaton field. Moreover the relation between the $(t,{\bar t})$-equations
and contact term algebras shed some light on its very
topological meaning.

Some open problems are suggested by our analysis that deserve a
more careful study. Next we briefly mention some of them.
A more mathematically understanding of the gravitational
picture we are using in this letter, will require to study
the equivariant cohomology \cite{Lo,O} which characterize
the physical states of the topological string theory obtained
by coupling to gravity ${\hat c} \! = \! 3$, $N \! = \! 2$
SCFT's, in the way prescribed by (1) and (2).

It is
well known that the contact term algebra of pure topological
gravity is equivalent to the Virasoro constraints for matrix
models \cite{DVV}, thus it is natural to ask what would be the analog for
the contact term algebras introduced in this letter. More
interesting will be to generalize our derivation of the $(t,{\bar
t})$-equations to the massive case.

{}From a more fundamental point of view and following
the line of thought initiated in references \cite{BCOV2,witten2}, we can
conceive the holomorphic anomaly as a way to define background
independence in string field theory. In this spirit the
topological theory defined by the contact term algebra seems to
indicate the type of dynamics would be necessary to add in order
to get background independence.

This work was partially supported by
grant PB 92-1092, the work of C.G. by Swiss National Science
Foundation and by OFES: contract number 93.0083, and the work of E.L. by
M.E.C. fellowship AP9134090983.

\end{document}